\documentclass[aps,pra,groupedaddress,twocolumn]{revtex4-1}

\usepackage{graphicx}
\usepackage{amsmath,amsfonts}

\def\myscale{1}

\usepackage{bm}
\newcommand{\Ai}{\operatorname{Ai}}

\graphicspath{{./fig/1/}{./fig/2/}{./fig/3/}{./fig/4/}}

\begin{document}

\title{Spatiotemporal diffraction-free pulsed beams in free-space of the Airy and Bessel type}

\author{Nikolaos K. Efremidis}
\email{nefrem@uoc.gr}
\affiliation{Department of Mathematics and Applied Mathematics, University of Crete, 70013 Heraklion, Crete, Greece}
\affiliation{Institute of Applied and Computational Mathematics, Foundation for Research and Technology - Hellas (FORTH), 70013 Heraklion, Crete, Greece}


\date{\today}

\begin{abstract}
We investigate the dynamics of spatiotemporal optical waves with one transverse dimension that are obtained as the intersections of the dispersion cone with a plane. We show that, by appropriate spectral excitations, the three different types of conic sections (elliptic, parabolic, and hyperbolic) can lead to optical waves of the Bessel, Airy, and modified Bessel type, respectively. We find closed form solutions that accurately describe the wave dynamics and unveil their fundamental properties. 
\end{abstract}

\maketitle

Diffraction-free pulses and beams are special classes of optical waves that can counteract diffraction and maintain their profile during propagation. Up to now a significant amount of effort has been devoted in the case of single frequency laser beams. Two are the main classes of beams with such a unique feature: The first is the Bessel beam~\cite{durni-prl1987,durni-josaa1987} that requires two-transverse directions and propagates along straight trajectories. Out of the infinite variants of the Bessel beam, representing different excitations of the dispersion circle, 
closed form expressions are also obtained in the case of Mathieu~\cite{vega-ol2000,cerda-ol2001,vega-oc2001} and Weber beams~\cite{bandr-ol2004,lopez-oe2005}. The second class is the Airy beam~\cite{sivil-ol2007,sivil-prl2007} that requires one transverse direction and bends along a parabolic trajectory as it propagates. The Airy beam is actually the only localized diffraction-free beam in one transverse direction~\cite{berry-ajp1979}. Generalizations of the Airy beam are obtained in higher dimensions either by utilizing the principle of superposition~\cite{sivil-ol2007} or by working in parabolic coordinates~\cite{bandr-ol2008,davis-oe2008}. 
In the non-paraxial limit a Bessel-type beam [obtained by accounting only the forward propagating plane waves (PWs) of a Bessel beam] is quasi diffraction-free~\cite{kamin-prl2012,zhang-ol2012}. By utilizing Babinet's principle it can be shown that all diffraction-free beams are able to self-reconstruct during propagation. Due to their unique properties, diffraction-free beams have found applications in microscopy~\cite{fahrb-np2010,jia-np2014,vette-nm2014}, laser filamentation~\cite{polyn-science2009}, beam autofocusing~\cite{efrem-ol2010,papaz-ol2011,zhang-ol2011}, ablation and micromachining~\cite{papaz-ol2011,mathi-apl2012}, and particle manipulation~\cite{baumg-np2008,zhang-ol2011} among others. 

On the other hand, diffraction-free spatiotemporal (ST) optical waves that contain a range of both temporal and spatial spectral components have been explored in the literature. These include Airy type bullets~\cite{chong-np2010}, X-waves~\cite{lu-tuf1992a,lu-tuf1992b,saari-prl1997}, Bessel pulse beams~\cite{shep-josaa2002}, and {different classes of Bessel X-waves~\cite{saari-lp1997,porra-pre2003,chris-ol2004}} among others. 
{Classes of spatiotemporally localized waves can be found by imposing a linear constraint between the propagation constant and the frequency~\cite{sonaj-ol1996,porra-pre2004}. In the nonlinear regime, due to nonlinear interactions X-waves can be spontaneously generated~\cite{facci-prl2006}.}
Recently there has been an effort in exploring the dispersion relation (DR) to synthesize waves in free-space with specific characteristics. In particular, in~\cite{konda-oe2016,parke-oe2016} waves, having the form of optical needles, with a spectral bandwidth that are associated with a constant longitudinal spatial frequency were constructed. Utilizing optical needles as building blocks, abrupt focusing and defocusing needles of light have been proposed~\cite{wong-acs2017}. Finally, in a recent work, it has been shown experimentally that the temporal degrees of freedom can be exploited to synthesize one-dimensional pulsed optical sheets that propagate self-similarly in free-space~\cite{konda-arxiv2017}. Such waves are associated with programmable conical correlations of the DR. 

In this work, we explore the possibilities of generating ST diffraction-free optical waves with one transverse dimension in free-space by applying specific constraints to the dispersion cone. These are obtained as the intersections of the conical DR with a plane and leads to elliptic, parabolic, and hyperbolic dispersion curves. In all three cases we have found that for specific spectral excitations and accounting for forward propagating PWs, optical pulsed beams described by incomplete Bessel and Airy integrals, and complete or incomplete modified Bessel integrals are derived. The resulting waves propagate in a diffraction-free fashion either forward or backward along the longitudinal direction with constant velocity. Interestingly the Airy-type wave, in addition to being diffraction-free, it also bends along a parabolic trajectory and exhibits a quasi self-similar profile in the transverse plane. The Bessel-type wave has a quasi-elliptic symmetry and, depending on the integer parameter $m$, a ST vorticity. Finally the modified Bessel-type solution takes the form of an X-wave with a focus in the center. Note that even in the incomplete cases, accurate closed form expressions are obtained by utilizing the complete versions of these integrals leading to Airy and Bessel functions. Experimentally, such waves can be made possible by using a grating to decompose the spectrum and then apply a phase pattern as demonstrated in~\cite{konda-arxiv2017}.

\begin{figure}
\centerline{
\includegraphics[width=\myscale\columnwidth]{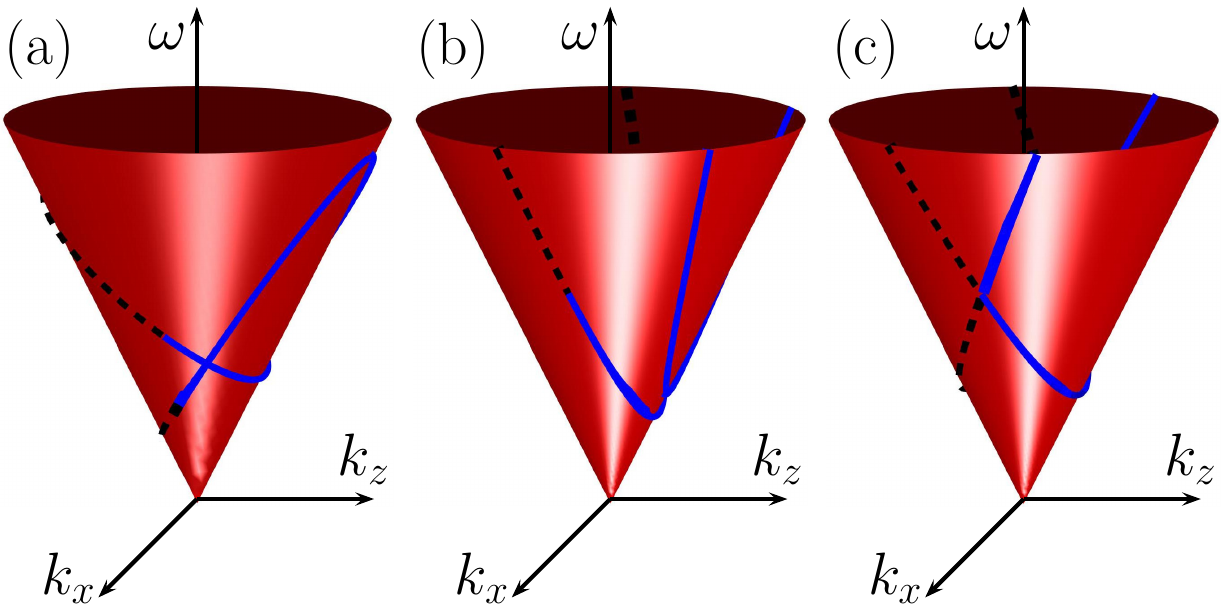}
}
\caption{(a) Elliptic (b) hyperbolic, and (c) parabolic intersections of the DR with a plane. The solid blue part of the curves corresponds to forward propagating PWs $k_z>0$ while the black dotted part is associated with $k_z<0$. \label{fig:1}}
\end{figure}
Let us start by considering an electric field $\bm E = \psi(x,z,t)\hat{\bm y}$ propagating in free-space that is uniform along its transverse polarization $y$-axis (and thus $\nabla\cdot\bm E=0$). 
The DR of such waves is
\begin{equation}
k_x^2+k_z^2=(\omega/c)^2,
\label{eq:dr}
\end{equation}
where $\omega>0$ is the optical frequency, $k_x$ and $k_z$ are the spatial frequencies along the $x$ and $z$ directions, and $c$ is the speed of light. The intersection of the inverted cone of Eq.~(\ref{eq:dr}) and the plane
$
k_z=s(\omega/v)-k_0
$
(for appropriate values of $k_0$) can give rise to, elliptic, hyperbolic, and parabolic curves as can be seen in Fig.~\ref{fig:1}. Specifically, if $s=\pm1$ and $v\ge0$ then elliptic curves have velocities $v<c$, hyperbolic curves have $v>c$ and parabolic curves exist for $v=c$. Selecting $x$ as the second transverse direction and $z$ as the longitudinal direction then a PW 
{
\begin{equation}
e^{i(k_xx+k_zz-\omega t)}
\label{eq:pw}
\end{equation}
}
is forward propagating if $k_z>0$. In this respect, a forward component wave (FCW) is any superposition of PWs with $k_z>0$. We graphically see in Fig.~\ref{fig:1} that in the parabolic and the elliptic cases any intersection of the dispersion cone with a plane consists of both forward and backward propagating PWs. On the other hand in the hyperbolic case, in addition to such mixed intersections, it is possible to construct intersections consisting exclusively of forward propagating PWs.

Before proceeding any further let as point out that in all the numerical simulations we use dimensionless coordinates. A simple method to go to dimensionless coordinates and avoid rewriting all the formulas it to scale the time variables with respect to $\tau$ and the space variables with respect to $c\tau$. The resulting expressions are dimensionless and identical to the original, except from the substitution $c\rightarrow1$. 

\begin{figure}[t]
\centerline{
\includegraphics[width=\myscale\columnwidth]{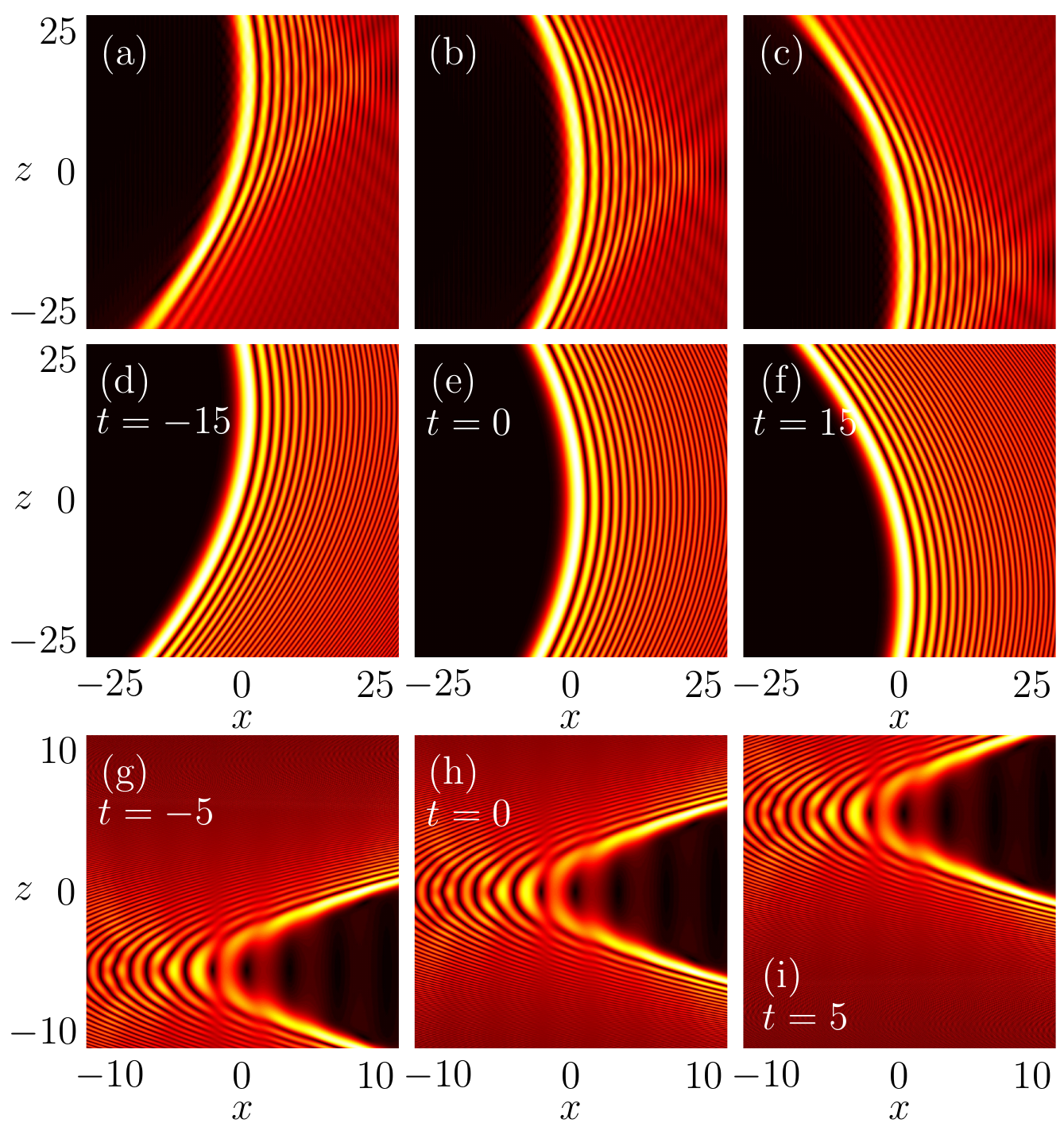}
}
\caption{{In the top and bottom rows we show the ST dynamics of the Airy-type FCW [Eq.~(\ref{eq:airyint})] in the case of a parabolic intersection of the DR with the plane and a cubic phase spectral excitation [Eq.~(\ref{eq:excitationAiry})]. In the first two rows we compare the FCW with the closed-form expression given by Eq.~(\ref{eq:airysol}). In the first two rows $l=-100$, $k_0=4$, $s=1$, whereas in the bottom row $l=3$, $k_0=1$, $s=-1$.\label{fig:2}}}
\end{figure}
Applying the constraint 
\begin{equation}
k_z=s(k_0-\omega/c)=s(\omega_0-\omega)/c
\end{equation}
to Eq.~(\ref{eq:dr}) results to a parabolic DR that can be expressed 
as a function of $k_x$ as
\[
\omega = c(k_x^2+k_0^2)/(2k_0),
\]
\[
k_z = s(k_0^2-k_x^2)/(2k_0).
\]
The two case $s=\pm1$ require separate treatment. For $s=1$ the condition $k_z>0$ results to $|k_x|<k_0$, whereas for $s=-1$ we obtain $|k_x|>k_0$. A generic interference of PWs {[Eq.~(\ref{eq:pw})]} with $k_z>0$ {[solid part of Fig.~\ref{fig:1}(c)]} results to the following FCW
\begin{equation}
\psi =
e^{\frac i2 k_0(sz-ct)}
\int_{s(k_0^2-k_x^2)>0} f(k_x)e^{ik_xx 
-ik_x^2\frac{(sz+ct)}{2k_0}} dk_x.
\label{eq:airyint}
\end{equation}
{Note that all the incomplete integrals encountered in this work are numerically integrated by using a simple trapezoid rule.}
By selecting a {Fourier space} cubic phase excitation 
\begin{equation}
f(k_x) =A\frac{(2k_0)^{2/3}}{2\pi|l|^{1/3}}
e^{il k_x^3/[3(2k_0)^2]},
\label{eq:excitationAiry}
\end{equation}
with $l$ real, 
Eq.~(\ref{eq:airysol}) becomes an Airy-type integral with incomplete integration limits. 
As we will see it is very instructive in terms of understanding the fundamental properties of the solution to relax the condition $k_z>0$ to $k_z\in\mathbb{R}$, and thus extend the limits of integration to the real line $k_x\in{\mathbb R}$.  Eq.~(\ref{eq:airyint}) then becomes  
\begin{equation}
\psi = Ae^{i\Psi_0}\Ai
\left(
\frac{xl-(sz+ct)^2}{|l|^{4/3}/(2k_0)^{2/3}}
\right),
\label{eq:airysol}
\end{equation}
with $\Ai$ being the Airy function, 
$
\Psi_0=k_0(sz-ct)/2+ 2k_0\xi (2\xi^2+3xl) /(3l^2),
$
and $\xi=sz+ct$. 
When $s=1$ in Eq.~(\ref{eq:airyint}) we integrate in the spatial frequency range $|k_x|<k_0$ which accounts for rays that bend up to a specific threshold. Thus the central part of the FCW is in excellent agreement with Eq.~(\ref{eq:airysol}) [{compare the first and the second rows of Fig.~\ref{fig:2}}]. Deviations start to appear at the tails of the caustic trajectory due to the fact that the FCW does not contain rays with strong bending that can follow the parabolic trajectory. The spectral bandwidth in this case is $[\omega_0/2,\omega_0]$. On the other hand, when $s=-1$, the bandwidth is $\omega>\omega_0$, and we integrate in the range $|k_x|>k_0$ that allows for the rays to bend only above a certain threshold. As a result, the strongly bending outer part of the FCW shows very good agreement with Eq.~(\ref{eq:airysol}) whereas the central part of the solution associated with weak bending is distorted [Fig.~\ref{fig:2}, bottom row]. Thus, in both cases, the exact Airy solution can be directly utilized to unveil the fundamental properties of the Airy-type FCW. From Eq.~(\ref{eq:airysol}) we see that the solution bends along a parabolic trajectory in the $x-z$ plane and has a constant acceleration in the $x-t$ plane. The parameter $l$ controls the bending, the acceleration, as well as the orientation (for positive and negative values) of the Airy wave along the $x$-direction. Furthermore the solution propagates backward ($s=1$) or forward ($s=-1$) with velocity $sc$. Note that both the FCW and Eq.~(\ref{eq:airysol}) are diffraction-free meaning that their amplitude profile remains invariant as time progresses. In addition, Eq.~(\ref{eq:airysol}) is also self-similar in the sense that independently of $z$ and $t$ it exhibits the same profile in the transverse $x$-direction. On the other hand, for example, the FCW shown in top row of Fig.~\ref{fig:2} is quasi self-similar up to the point where the rays can bend along the parabolic trajectory. Note that $k_0$ not only controls the bandwidth, but also the spatial frequency of the oscillations of the Airy lobes along the $x$-direction.

\begin{figure}
\centerline{
\includegraphics[width=\myscale\columnwidth]{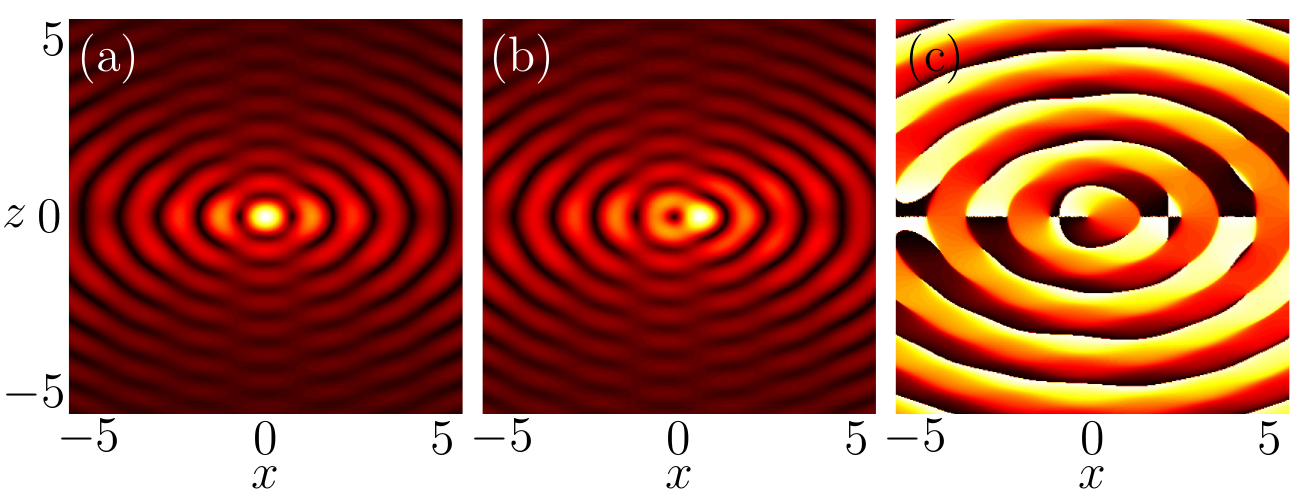}
}
\caption{Profiles of the Bessel-type FCWs [Eq.~(\ref{eq:besselint})] in the case of an elliptic conic section and a spectral excitation given by Eq.~(\ref{eq:excitationBessel}). In (a) the amplitude is shown for $m=0$, whereas in (b)-(c) both amplitude and the phase are depicted for $m=1$. In the phase pattern we have excluded the term $\Psi_0$. Parameters are $s=1$, $k_0=2$, $v=0.8$. \label{fig:3}}
\end{figure}
An elliptic conic section is obtained by intersecting the inverted dispersion cone with the plane
\[
k_z=s(\omega/v-k_0)=s(\omega -\omega_0)/v,
\]
with $0<v<c$, $s=\pm1$, and $k_0>0$. We can parametrically express the conic section in terms of $\theta$ as
\[
k_x =k_0(v/\Delta)\sin\theta,
\]
\[
k_z=sk_0(v/\Delta)^2[1+(c/v)\cos\theta],
\]
\[
\omega = \omega_0(c/\Delta)^2 [1+ (v/c)\cos\theta],
\]
where $\Delta = \sqrt{c^2-v^2}$.
The condition for forward propagating PWs {$k_z>0$ [solid part of the curve in Fig.~\ref{fig:1}(a)]} translates to 
$|\theta|<\pi-\arccos(v/c)$ for $s=1$ and to 
$\pi-\arccos(v/c)<|\theta|<\pi$ for $s=-1$. A generic interference of such PWs leads to the FCW 
\begin{equation}
\psi =
e^{i\Psi_0}
\!\int_{k_z>0}\!
 f(k_x) e^{
i\frac{k_0cv}{\Delta^2}
\left[
\frac{\Delta x }{c}\sin\theta +
(s z-vt) \cos\theta
\right]
}d\theta,
\label{eq:besselint}
\end{equation}
with
$ 
\Psi_0 = -\omega_0t+k_0v^2(sz-vt)/\Delta^2. 
$
By selecting the following spectral excitation
\begin{equation}
f(\theta) = Ae^{-im\theta}/(2\pi)
\label{eq:excitationBessel}
\end{equation}
with integer $m$ the integral becomes of the Bessel-type with incomplete limits of integration. Relaxing the constraint $k_z>0$ to $\theta\in[-\pi,\pi]$ leads to a closed-form expression in terms of the Bessel functions $J_m$ as
\begin{equation}
\psi = A e^{i\Psi_0+im\phi}
J_m\left[
\frac{k_0cv}{\Delta^2}
\left(
\frac{\Delta^2x^2}{c^2}+(sz-vt)^2
\right)^{1/2}
\right]
\label{eq:besselsol}
\end{equation}
with
\[
\tan\phi = c(sz-vt)/(\Delta x).
\]
When the integration parameter $\theta$ with $k_z(\theta)>0$ covers most of the ellipse [see for example the ellipse of Fig.~\ref{fig:1}(a) with a positive inclination in the $k_z-\omega$ plane] then the resulting Bessel-type FCW is in good agreement with the exact solution of Eq.~(\ref{eq:besselsol}). This is possible when $v$ is relatively close to $c$ and $s=1$. 
These waves propagate in a diffraction-free fashion in the $z-t$ plane with (positive or negative) velocity $s v$ and have quasi-elliptic Bessel-type symmetry with vorticity $m$ in the $x-z$ and the $x-t$ planes. 
In Fig.~\ref{fig:3}(a) $m=0$ and we see that the elliptic symmetry is slightly broken but the amplitude of the FCW still maintains its parity along its $x$ and $z$ axes and is in good agreement with Eq.~(\ref{eq:besselsol}). 
For $m=1$, in Fig.~\ref{fig:3}(b)-(c) we observe that the parity along the $x$-axis is also broken but the parity along the $z$-axis is maintained. In comparison to the Bessel solution of Eq.~(\ref{eq:besselsol}) a higher intensity lobe appears on the right side of the first Bessel-type ring. Importantly, the vortex phase structure is maintained with the location of the vortex core being shifted towards the negative $x$-axis. Note that in Fig.~\ref{fig:3}(c) in the computation of the phase we have excluded the term $\Psi_0$ that does not contribute to the generation of the vortex but complicates the visualization of the vortex phase pattern. The spectral bandwidth of the FCW lies in the region between $\omega_0$ and $\omega_0(c/(c-sv))^{1/2}$ (the ordering of these two values depends on $s$). In Eq.~(\ref{eq:besselsol}) $k_0$ controls the spacing between the concentric quasi-elliptic rings whereas $\Delta/c$ is the eccentricity of the ellipse.

\begin{figure}
\centerline{
\includegraphics[width=0.8\columnwidth]{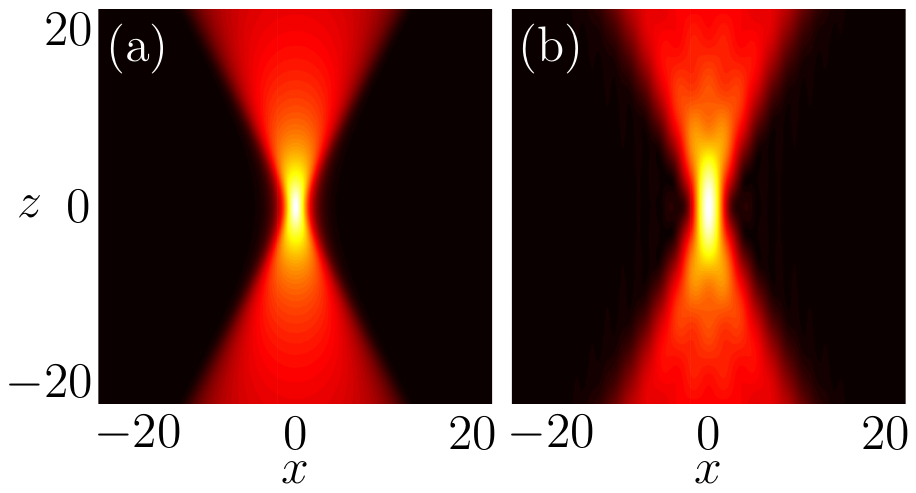}
}
\caption{Amplitude profile of the FCWs in the case of a hyperbolic conic section and a spectral apodization given by Eq.~(\ref{eq:apodization}). Parameters are $v=2$, $t=0$, $t_0=1$ while in (a) $s=1$, $k_0=1$, and in (b) $s=-1$, $k_0=-1$.\label{fig:4}}
\end{figure}
Finally, classes of diffraction-free waves exist when the conic section of the plane with the DR is a hyperbolic curve. We express the equation for the plane as 
\[
k_z = s(k_0+\omega/v)= s(\omega+\omega_0)/v
\]
with $s=\pm1$, $k_0\in\mathbb{R}$, and $v>c$. We would like to point out that, as it can be seen in Fig.~\ref{fig:1}(b), in this case the conic section can consist of either both forward and backward propagating PWs or only forward propagating PWs. The hyperbolic curve can be parametrically written as
\[
k_x =k_0(v/\Delta)\sinh\sigma,
\]
\[
k_z=
s(v/\Delta)^2
[k_0+|k_0|(c/v)\cosh\sigma],
\]
\[
\omega = |\omega_0|(c/\Delta)^2
[1+(v/c)\cosh\sigma],
\]
for $\Delta=\sqrt{v^2-c^2}$ and real values of $\sigma$. A generic interference of such FCW is  given by 
\begin{equation}
\psi = e^{i\Psi_0}
\!\!\int_{k_z>0} \!
f(\sigma)
e^{
\frac{i|k_0|cv}{\Delta^2}
[
\frac{\Delta x }{c}\sinh\sigma +
(sz-vt) \cosh\sigma
]}
d\sigma
\label{eq:besselkint}
\end{equation}
with
$
\Psi_0 = 
\omega_0t +  
k_0v^2(s z-vt)/\Delta^2.
$
and $\sigma$ satisfying the condition $k_z(\sigma)>0$. 
In order to avoid singularities in the resulting solution, in a similar fashion with~\cite{parke-oe2016}, we select the spectral apodization function
\begin{equation}
f(\sigma) = 2A\exp\left[-|k_0|ct_0 \left(\frac v \Delta\right)^2\cosh\sigma\right].
\label{eq:apodization}
\end{equation}
By substituting Eq.~(\ref{eq:apodization}) to Eq.~(\ref{eq:besselkint}) and extending (if necessary) the limits of integration to the real line we obtain
\begin{equation}
\psi = Ae^{i\Psi_0}
K_0\left[
\frac{|k_0|cv}{\Delta^2}
\left(
\frac{\Delta^2x^2}{c^2}-(sz-v(t-it_0))^2
\right)^{1/2}
\right],
\label{eq:besselksol}
\end{equation}
where $K_0$ is a modified Bessel function. As can be seen in Fig.~\ref{fig:4}, the solution takes the form of a ST $X$-wave in the $x-t$ and the $x-z$ planes 
that propagates along the longitudinal $z$-direction with velocity $sv$. In Fig.~\ref{fig:4}(a) the conic section consists of only forward propagating waves and thus Eq.~(\ref{eq:besselksol}) is the exact solution of Eq.~(\ref{eq:besselkint}). On the other hand, in Fig.~\ref{fig:4}(b) the conic section contains both forward and backward propagating components. However, the comparison between the FCW as described by the incomplete modified Bessel-type integral of Eq.~(\ref{eq:besselkint}) and Eq.~(\ref{eq:besselksol}) is very good. As in the previously described cases, all the FCWs of Eq.~(\ref{eq:besselkint}) are diffraction free. Note that in the case $k_0>0$ and $s>0$ by taking the limit $v\rightarrow\infty$ the needle wave {with constant $k_z$} is obtained in~\cite{parke-oe2016} is recovered. 

{For completeness, let us consider the case of straight line intersections with the DR. Without loss of generality we assume $k_z=\omega/c$, $k_x=0$ leading to $\psi = F(z/c-t)= (1/(2\pi))\int_{-\infty}^\infty f(\omega)e^{i\omega(z/c-t)}d\omega$. The condition for forward propagating plane-waves is satisfied by setting $f(\omega)=0$ for $\omega<0$.}

In conclusion, we have found classes of ST optical waves with one transverse dimension that propagate in a diffraction-free fashion. The solutions are obtained as conic sections of the DR for particular spectral excitation and take the form of complete/incomplete modified Bessel, and incomplete Airy and Bessel functions. Such solutions can be experimentally realized by applying a spatiotermporal phase pattern to the optical wave~\cite{konda-arxiv2017}.

\noindent{\textbf{Funding.}} 
Erasmus Mundus NANOPHI Project (2013- 5659/002-001). 

\newcommand{\noopsort[1]}{} \newcommand{\singleletter}[1]{#1}

\end{document}